\documentclass[12pt]{article}

\input epsf
\usepackage{amssymb,amsmath,subfigure,float}
\usepackage[matrix,arrow,curve]{xy}
\usepackage{cite}
\usepackage{multirow}
\usepackage{graphicx,color}
\usepackage[debug,pageanchor=false]{hyperref}
\definecolor{link}{rgb}{.8,.15,.1}
\hypersetup{colorlinks=true,linkcolor=link,citecolor=link,urlcolor=link,linktocpage}

\usepackage{pgf}
\usepackage{tikz}
\usepackage[utf8]{inputenc}
\usetikzlibrary{arrows,automata}
\usetikzlibrary{positioning}

\tikzset{
    state/.style={
           rectangle,
           rounded corners,
           draw=black, very thick,
           minimum height=2em,
           inner sep=2pt,
           text centered,
           },
}

\makeatletter
\@addtoreset{equation}{section}
\makeatother

\begin{document}

\begin{titlepage}

\begin{center}

\vskip .3in \noindent

{\Large \bf{6d $\rightarrow$ 5d $\rightarrow$ 4d reduction of BPS attractors\\ in flat gauged supergravities \\\vspace{.2cm}}}

\bigskip

	Kiril Hristov, Andrea Rota\\

       \bigskip
		 Dipartimento di Fisica, Universit\`a di Milano--Bicocca, I-20126 Milano, Italy\\
       and\\
       INFN, sezione di Milano--Bicocca,
       I-20126 Milano, Italy

       \vskip .5in
           
       {\bf Abstract }
        \vskip .1in        
        \end{center}         
       
	Via a series of Kaluza-Klein (KK) and Scherk-Schwarz (SS) compactifications we relate BPS attractors and their complete (in general non-BPS) flows to a Minkowski vacuum in gauged supergravities with vanishing scalar potential in 4, 5, and 6 dimensions. This way we can look at a class of extremal non-BPS black holes and strings from IIB string theory viewpoint, keeping 4 supercharges on the horizon. Our results imply the existence of a dual 2d $N = (0,2)$ superconformal field theory (SCFT) that originates from a parent $N=(4,4)$ theory living on a D1-D5 system. 
	
	This is achieved starting from the BPS black string in 6d with an AdS$_3 \times$S$^3$ attractor and taking two different routes to arrive at a 1/2 BPS AdS$_2 \times$S$^2$ attractor of a non-BPS black hole in 4d $N=2$ flat gauged supergravity. The two inequivalent routes interchange the order of KK reduction on AdS$_3$ and SS reduction on S$^3$. We also find the commutator between the two operations after performing a duality transformation: on the level of the theory the result is the exchange of electric with magnetic gaugings; on the level of the solution we find a flip of the quartic invariant $I_4$ to $-I_4$.


\noindent

\vfill
\eject

\end{titlepage}

\section{Introduction and summary of results} 
\label{sec:intro}
Dimensional reduction between black hole solutions in string theory has led to important developments of the field \cite{Breckenridge:1996is,LozanoTellechea:2002pn,Gaiotto:2005gf,Gaiotto:2005xt,Behrndt:2005he,Banerjee:2011ts}. The relation between supersymmetric black (st)rings and black holes in ungauged supergravities in 6d/5d and 5d/4d was crucial for the microscopic understanding of black hole entropy \cite{Strominger:1996sh,Maldacena:1997de} and has therefore given us a tool to look into the quantum regime of black hole physics. Here and in a companion paper \cite{Hristov:2014eza} we explore similar relations between supersymmetric black objects in 4, 5, and 6 dimensions, this time in gauged supergravity. In particular here we look at the dimensional reduction for gauged theories with a vanishing scalar potential, such that one has the same bosonic lagrangian as in ungauged supergravity with asymptotically flat black hole solutions. This is interesting to do because the gauged theory has different set of BPS vacua with respect to the ungauged one, even if the full spectrum of bosonic solutions is the same \cite{Hristov:2012nu}. We find that one needs to use Scherk-Schwarz (SS) instead of Klauza-Klein (KK) reduction in order to preserve some fraction of supersymmetry on the horizon in the lower-dimensional gauged theory. In this way we find a string theory interpretation to the BPS attractors of non-BPS black holes \cite{Hristov:2012nu} and strings \cite{Hristov:2013xza} and thus we can understand better their field theory duals.

It has been understood that SS reduction of a theory of ungauged supergravity leads to a gauged supergravity with a vanishing scalar potential in the lower dimension \cite{Bergshoeff:1997mg,Lavrinenko:1997qa,Andrianopoli:2002mf,Andrianopoli:2004im,Andrianopoli:2004xu,Andrianopoli:2005jv,Dall'Agata:2005ff,Hull:2006tp,Looyestijn:2010pb}. We exploit this fact\footnote{Strictly speaking, there is a distinction between two classes of SS reductions: reductions over a circle with duality twist and the case of twisted tori or twistings of other manifolds, as explained in detail in \cite{Hull:2006tp}. Here we perform a reduction with a duality twist over a circle, to be defined more precisely in the next section.} and connect the BPS attractor of black strings in 6d, AdS$_3 \times$S$^3$, with BPS black string and black hole attractors in 5d and 4d gauged supergravities. Depending on the choice of signs for the charges and gauge coupling constant in the SS reduction, we can end up with 1/2 BPS attractors in gauged $N=2$ or fully BPS attractors in ungauged $N=2$ supergravity\footnote{Here and in the following sections we mostly focus our discussion on the 1/2 BPS attractors coming from the SS reduction as the fully BPS ones in ungauged supergravity are already well-known and understood.} (in the limit where the SS reduction becomes KK). The 1/2 BPS attractors are the near-horizon geometries of extremal non-BPS black holes and strings in 4 and 5 dimensions \cite{Hristov:2012nu,Hristov:2013xza} (see \cite{LopesCardoso:2007ky,Gimon:2007mh,Gaiotto:2007ag,Goldstein:2008fq,Andrianopoli:2009je,Perz:2008kh,Bena:2009en,Bossard:2012xsa} and references therein for extensive results on extremal non-BPS black holes and their horizons). We find preserved supersymmetry on the horizon only in case of SS reduction on the internal space S$^3$ and KK reduction on the AdS$_3$. Since the starting 6d attractor can be seen as a string theory background of type IIB on K3 or T$^4$ corresponding to a D1-D5 system, the resulting 4d attractors also have a string theory interpretation.

As shown on the flow chart below, we take the starting geometry AdS$_3 \times$S$^3$ and follow two inequivalent paths down to AdS$_2 \times$S$^2$ in 4d, both of which can preserve supersymmetry. Path I is to first KK reduce along the circle inside AdS$_3$ and only then perform the SS reduction from S$^3$ to S$^2$, while path II is the inverse - first SS on S$_3$ and then KK on AdS$_3$. The two inequivalent paths from 6d to 4d give the same bosonic lagrangian but different solutions, which we show to be related by a duality transformation up to a flip of sign in the quartic invariant I$_4$. The same duality transformation is also a symmetry of the bosonic lagrangian, while in the fermonic sector it interchanges electric and magnetic gaugings, leading to a nontrivial commutator between the two reduction paths.

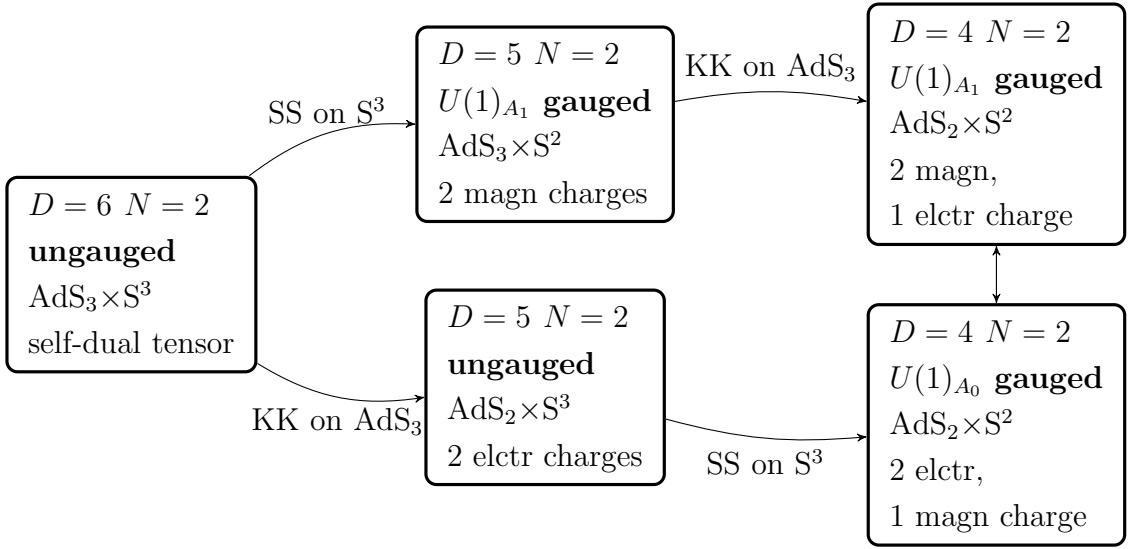
\begin{tikzpicture}[->,>=stealth']\label{fig1}

 \node[state] (QUERY) 
 {\begin{tabular}{l}
  \textbf{$D=6$ $N=2$}\\
  \textbf{ungauged}\\
  AdS$_3 \times$S$^3$\\
  self-dual tensor\\

 \end{tabular}};
  
 \node[state,    	
  yshift=2cm, 		
  right of=QUERY, 	
  node distance=5.5cm, 	
  anchor=center] (ACK) 	
 {%
 \begin{tabular}{l} 	
  \textbf{$D=5$ $N=2$}\\
  \textbf{$U(1)_{A_1}$ gauged}\\
  AdS$_3 \times$S$^2$\\
  2 magn charges
 \end{tabular}
 };
 
 \node[state,
  below of=ACK,
  node distance=1.5cm,
  yshift=-2cm,
  anchor=center] (QUERYREP) 
 {%
\begin{tabular}{l} 	
  \textbf{$D=5$ $N=2$}\\
  \textbf{ungauged}\\
  AdS$_2 \times$S$^3$\\
  2 elctr charges
 \end{tabular}
 };

 \node[state,
  right of=ACK,
  node distance=6cm,
  anchor=center] (EPC) 
 {%
\begin{tabular}{l} 	
  \textbf{$D=4$ $N=2$}\\
  \textbf{$U(1)_{A_1}$ gauged}\\
  AdS$_2 \times$S$^2$\\
  2 magn,\\ 1 elctr charge
 \end{tabular}
 };

  \node[state,
  below of=EPC,
  node distance=4cm,
  anchor=center] (BELEPC) 
 {%
\begin{tabular}{l} 	
  \textbf{$D=4$ $N=2$}\\
  \textbf{$U(1)_{A_0}$ gauged}\\
  AdS$_2 \times$S$^2$\\
  2 elctr,\\ 1 magn charge
 \end{tabular}
 };
 
 \path (QUERY) 	edge[bend left=20]  node[anchor=south,above]{SS on S$^3$} (ACK)
                                    
 (QUERY)     	edge[bend right=20] node[anchor=south,below]{KK on AdS$_3$} (QUERYREP)
 (ACK)       	edge[bend left=10]  node[anchor=south,above]{KK on AdS$_3$}  (EPC)
 (QUERYREP)	edge[bend right=10] node[anchor=south,below]{SS on S$^3$} (BELEPC)
 (BELEPC)       edge                                          (EPC)
 (EPC)       	edge                                          (BELEPC);

\end{tikzpicture}\\

The fact that we find preserved supersymmetry in our analysis is a crucial point, which allows us to claim that the dual field theories on the horizon of the black strings/holes are also supersymmetric. This means that we have a supersymmetric version of the AdS/CFT correspondence and can trust the black hole microstate counting even if the full black hole/string geometry is non-BPS. We confirm this by verifying that the central charge and macroscopic entropy of the non-BPS black string and black hole, respectively, can be recovered correctly from a dual field theory description based on the D1-D5 field theory\footnote{Note that the IR limit of the D1-D5 field theory is a $N=(4,4)$ 2d SCFT and is dual to the black string attractor in 6d with a symmetry algebra $SU(1,1|2) \times SU(1,1|2)$. In our considerations we already start with $N=2$ supergravity in 6d, where we only have 8 supercharges and therefore a symmetry group $SU(1,1|2) \times SU(1,1) \times SU(2)$. This does not change the fact that there secretly are more fermionic symmetries in 6d, which are not present anymore after the reduction. Apart from this subtlety in 6d that should be kept in mind, the other supersymmetry groups cited in this paper are strictly valid and cannot be extended.} \cite{Strominger:1996sh} (see also section 5.3 of \cite{Aharony:1999ti} and references therein for more details).  

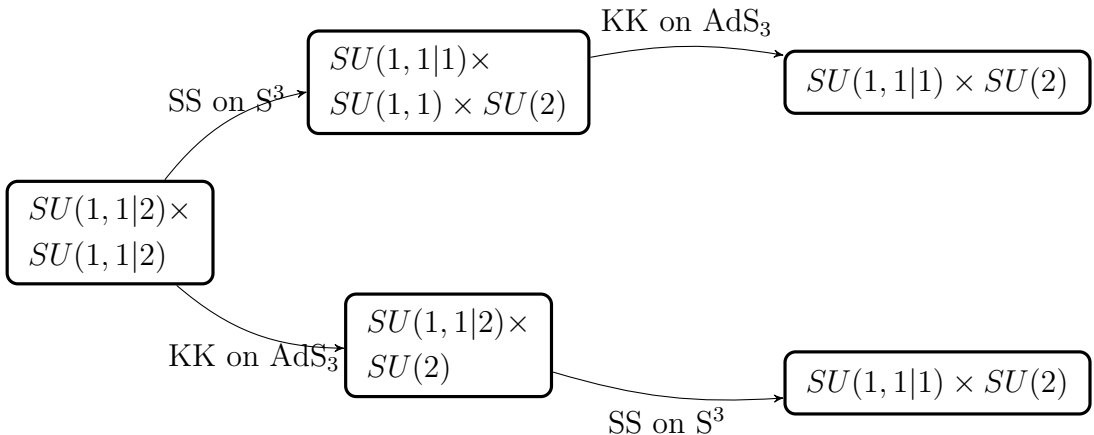
\begin{tikzpicture}[->,>=stealth']\label{fig2}

 \node[state] (QUERY) 
 {\begin{tabular}{l}

  $SU(1,1|2) \times$\\
  $SU(1,1|2)$\\
  
 \end{tabular}};
  
 \node[state,    	
  yshift=2cm, 		
  right of=QUERY, 	
  node distance=4.5cm, 	
  anchor=center] (ACK) 	
 {%
 \begin{tabular}{l} 	
  $SU(1,1|1) \times$\\
  $SU(1,1)\times SU(2)$\\
 \end{tabular}
 };
 
 \node[state,
  below of=ACK,
  node distance=1.5cm,
  yshift=-2cm,
  anchor=center] (QUERYREP) 
 {%
\begin{tabular}{l} 	
  $SU(1,1|2) \times$\\
  $SU(2)$\\
 \end{tabular}
 };

 \node[state,
  right of=ACK,
  node distance=6.5cm,
  anchor=center] (EPC) 
 {%
\begin{tabular}{l} 	
  $SU(1,1|1) \times SU(2)$\\
 \end{tabular}
 };

  \node[state,
  below of=EPC,
  node distance=4cm,
  anchor=center] (BELEPC) 
 {%
\begin{tabular}{l} 	
 $SU(1,1|1) \times SU(2)$\\
 \end{tabular}
 };
 
 \path (QUERY) 	edge[bend left=20]  node[anchor=south,above]{SS on S$^3$} (ACK)
                                    
 (QUERY)     	edge[bend right=20] node[anchor=south,below]{KK on AdS$_3$} (QUERYREP)
 (ACK)       	edge[bend left=10]  node[anchor=south,above]{KK on AdS$_3$}  (EPC)
 (QUERYREP)	edge[bend right=10] node[anchor=south,below]{SS on S$^3$} (BELEPC);

\end{tikzpicture}
  
Note that the SS reduction of the full black string geometry in general does break fully supersymmetry, and BPS-ness is restored only on the black string/hole horizon as shown in the symmetry algebras above. Thus the corresponding ``SS reduction'' operation that one needs to perform on the D1-D5 branes also breaks supersymmetry, and the resulting theory has a restored $N=(0,2)$ superconformal symmetry only at its IR fixed point. We show that the most naive expectation for the ``SS reduction'' operation on the D1-D5 theory, namely that states charged under the original $SU(2)_R$ are projected out, agrees with the value of the central charge from the AdS/CFT dictionary at leading order. It remains an open question whether one can define more precisely the exact operation of ``SS reduction'' on the D1-D5 system and perform other nontrivial checks on it.  
  
The plan of the paper is as follows. First we make some general comments and explain conceptually the difference between SS and KK reductions in section \ref{sec:ss-kk}. We thus gain some more intuition before proceeding to the supergravity details in sections \ref{sec:6to5} and \ref{sec:5to4}, where we explictly construct the reduction from 6d to 5d and from 5d to 4d along the two different paths outlined above. Then in section \ref{sec:commutator} we construct an explicit duality transformation that relates the two final solutions and thus derive the commutator between KK-SS and SS-KK. We finish the main part of the paper in section \ref{sec:dualCFT} with the dual conformal field theory picture. The fact that we manage to relate the different attractor geometries gives a suggestion how to derive the dual field theories starting from the original D1-D5 theory, on which we make some more general comments and finally conclude with section \ref{sec:conclusion}.

\section{Scherk-Schwarz vs. Kaluza-Klein reduction} 
\label{sec:ss-kk}
There are plenty of comprehensive references discussing in detail KK and SS reductions (or duality twists) \cite{Breckenridge:1996is,Andrianopoli:2004im,Andrianopoli:2004xu,Andrianopoli:2005jv,Looyestijn:2010pb}, showing explicitly the relation between the supergravities we are interested in. We can however sketch the basic mechanism, which for our purposes is very simple. One can generate Fayet-Iliopoulos (FI) terms in the lower dimensional theory simply by allowing for a reduction ansatz for the gravitino of the form 
\begin{equation}
 \label{SSgravitino}
 \Psi_{\hat{\mu}} (\hat{x}) = e^{i g y} \Psi_{\hat{\mu}} (x)\ , 
\end{equation}
where $\hat{x} = {x, y}$ are the original coordinates, and g is an arbitrary constant. This choice is always allowed in the supergravity theories we consider since the R-symmetry group contains a $U(1)$ subgroup and therefore the phase in \eqref{SSgravitino} is a symmetry of the lagrangian. For a nonvanishing $g$ this choice breaks the R-symmetry group (in our case $SU(2)_R$) to a $U(1)$ subgroup. Note that here one can easily get back to the standard KK reduction by taking the limit $g=0$. To see how one gets additional FI terms in the lower dimensional action it is enough to consider the kinetic term for the gravitino
\begin{equation}
 \label{kinetic_gravitino}
 \mathcal{L}_{d+1} = \bar{\Psi}_{\hat{\mu}} \gamma^{\hat{\mu} \hat{\nu} \hat{\rho}} \partial_{\hat{\nu}} \Psi_{\hat{\rho}}\ ,
\end{equation}
together with the standard metric decomposition 
\begin{equation}\label{KKmetric}
  {\rm d} s^2_{d+1} = e^{-2 \phi} {\rm d} s^2_d + e^{2 (d-2) \phi} ({\rm d} y + A_{\mu} {\rm d} x^{\mu} )^2\ ,
\end{equation}
chosen such that one goes from Einstein frame in $(d+1)$ dimensions to the Einstein frame in $d$ dimensions.
The nonvanishing components of the vielbein and its inverse are given by
\begin{equation}
 \label{vielbein}
 \hat{e}_{\mu}^a = e^{-\phi} e_{\mu}^a,\ \hat{e}_{\mu}^{10} = e^{(d-2) \phi} A_{\mu},\ \hat{e}_y^{10} = e^{(d-2) \phi}; \quad \hat{e}_a^{\mu} = e^{\phi} e_a^{\mu}, \ \hat{e}_{10}^y = e^{-(d-2) \phi},\ \hat{e}_a^y = -e^{\phi} A_a, 
\end{equation}
where $e_{\mu}^a$ and $e_a^{\mu}$ are the lower dimensional vielbein and its inverse, and the label $10$ signifies the flat index of the coordinate that is reduced upon. 

The lower dimensional gravitino terms that one finds from \eqref{kinetic_gravitino} become\footnote{We ignore lower-dimensional half-spin fields that result from the reduction of the gravitino, i.e.\ terms that include $\Psi_y$.}
\begin{equation}
 \label{reduced_gravitino}
 \mathcal{L}_d = \bar{\Psi}_{\mu} \gamma^{\mu \nu \rho} \partial_{\nu} \Psi_{\rho} - i g \bar{\Psi}_{\mu} \gamma^{\mu \rho} (e^{(d-2) \phi} \gamma^{10} - e^{\phi} A_{\nu} \gamma^{\nu}) \Psi_{\rho} + ...  
\end{equation}
One can see that the first term in the brackets gives a (scalar dependent) mass for the gravitino, while the second term can be recombined into a covariant derivative $D_{\mu} \Psi_{\nu}$ in the lower dimensional theory, such that the gravitino carries charge proportional to $g$ under the Kaluza-Klein gauge field $A_{\mu}$. These two terms make the difference between gauged and ungauged supergravity in the simplest models and one refers to such theories as FI gauged ones. It is clear that both terms vanish when $g=0$, leading back to the standard ungauged supergravity from KK reductions. A crucial point for us here is that in the SS reduction we have defined above the constant $g$ is arbitrary and not governed by any higher dimensional dynamics. We are therefore free to fix it to any value, which is important in the next sections when we discuss how the explicit solutions under consideration preserve supersymmetry. Anticipating our analysis there we already note that since the gravitini carry an electric charge proportional to $g$, the magnetic charge of the solutions we find carried by the KK gauge field is quantized in inverse units of $g$. 

The explicit supergravity models we consider in the next section of course posses a larger number of fields, both bosonic and fermionic. However we only perform SS reduction on the gravitini, keeping the remaining fields uncharged under the KK gauge field. This means that our bosonic action is always that of ungauged supergravity, since all bosonic fields are KK reduced. Therefore the gauging of the gravitino does not produce any cosmological constant or scalar potential and we remain within the class of ``flat'' gauged supergravities as promised in the introduction. 

The exact details of the SS reductions from 6d to 5d \cite{Breckenridge:1996is,Andrianopoli:2004xu} and 5d to 4d \cite{Andrianopoli:2004im,Behrndt:2005he,Looyestijn:2010pb} are already known and we directly use them in the following sections.

\section{6D to 5D} 
\label{sec:6to5}

We start from ungauged $N = 2$ supergravity in six dimensions\footnote{We could have started with more general $N=4$ supergravity, since the black strings of Strominger-Vafa \cite{Strominger:1996sh} are to be found there. The theory we consider is still a truncation of the $N=4$ so the reductions we perform in this section can be also thought of as starting from ungauged $N=4$ in 6d and leading to flat gauged $N=4$ in 5d. However, due to the $N=2$ truncation we take we miss some other possible reductions of the larger $N=4$ theory, i.e.\ the ones with nontrivial tensor reduction ansatz performed in \cite{Villadoro:2004ci}.}. The $N=2$ gravity multiplet contains two Majorana-Weyl gravitini and a self dual antisymmetric tensor field\cite{Nishino:1984gk}:
\begin{equation}
\hat{G}_{\hat{M}\hat{N}} \ , \ \Psi^{A}_{\hat{M}} \ , \ B_{\hat{M}\hat{N}}, 
\end{equation}
where $A=1,2$ runs over the fudamental of the R symmetry group $Sp(1)$. The bosonic lagrangian has the following simple form\cite{Marcus:1982yu,Nishino:1986dc,Cariglia:2004kk}:
\begin{equation}
\mathcal{L}_6 = \sqrt{\hat{G}} \left(-\frac{1}{2}R +\frac{1}{6}H_{\hat{M}\hat{N}\hat{P}}H^{\hat{M}\hat{N}\hat{P}}  \right),
\end{equation}
where $H=dB$. Our interest in this theory is related to the presence of an AdS$_3\times$S$^3$ backgound, which is fully BPS and corresponds to the near horizon limit of the self dual string soliton \cite{Gibbons:1994vm,Gutowski:2003rg}:
\begin{equation}
ds^2_6 = L^2 ({\rm d} s^2_{AdS_3}+{\rm d} s^2_{S^3})\ ,
\end{equation}
where:
\begin{equation}\label{ads_j}
 {\rm d} s^2_{AdS_3} = \frac{1}{4} \left(-\cosh^2 \beta\ {\rm d} \alpha^2 + {\rm d} \beta^2 + \left( \frac{{\rm d} \gamma}{J}+\sinh \beta\ {\rm d} \alpha \right)^2 \right)\ ,
\end{equation}
\begin{equation}\label{s_k}
  {\rm d} s^2_{S^3} = \frac{1}{4} \left(\sin^2 \theta\ {\rm d} \phi^2 + {\rm d} \theta^2 + \left( \frac{{\rm d} \psi}{K}- \sigma \cos \theta\ {\rm d} \phi \right)^2 \right)\ .
\end{equation}

More precisely this metric corresponds to a near-horizon geometry of extremal BTZ with a parameter $\rho_+ = 1/2J$ in the standard notation and a sphere with coordinate\footnote{We have rescaled the metric for the sphere with $K$ and then changed the period of the coordinate $\psi$ accordingly. This means the sphere is left untouched, but we have used slightly odd coordinate choice, which allows us to find more general Kaluza-Klein charges. Alternatively, we could leave the metric in its usual form but choose a slight generalization of the reduction ansatz, as done in \cite{Behrndt:2005he}. These two choices are completely equivalent and do not lead to real change in physics in either dimension.} $\psi/ K \in [0, 4 \pi)$. We leave the parameter $\sigma =  \{ + , - \} $ unspecified since it gives the sign of a magnetic charge in 5/4d. This is important when we discuss the supersymmetry properties after the reduction.

The BPS string in 6d further has a nonvanishing self-dual tensor field, given by
\begin{equation}
 H = L^2 \left(Vol (AdS_3) + Vol (S^3) \right)\ .
\end{equation}
For the dimensional reduction of the fields to five dimensions we take the same field ansatz for both KK and SS reductions: 
\begin{equation}\label{redrules65}
 {\rm d} s^2_6 = z^{-1} {\rm d} s^2_5 + z^3 ({\rm d} x_6 + L_6 ^{-2} A^1)^2\ \ ,  \   \ B_{M6}=L_6 A^2_M,
\end{equation}
where $L_6$ is the lenght of the circle upon which we reduce. The presence of powers of $L_6$ in the reduction ansatz is needed in order to get the proper normalization in five dimensions. The resulting lower-dimensional fields can be organized into an $N=2$ 5d gravity multiplet containing $\{ G_{M N}, A^1_M \} $ plus a vector multiplet containing $\{A^2_M, z \}$ as already shown in \cite{Andrianopoli:2004xu} and the bosonic lagrangian reads:
\begin{equation}\label{5dlagr}
\mathcal{L}_5 = \sqrt{G} \left(-\frac{1}{2}R +\frac{3}{8 z^2} \partial_M z \partial^M z + \frac{1}{8 z^2} F^1_{MN} F^{1,MN} + \frac{z}{4} F^2_{MN} F^{2,MN} \right) + \frac{1}{12} C_{IJK} F^{I} \wedge F^{J} \wedge A^{K}\ ,
\end{equation}
\noindent with nonvanishing $C_{122}=2$ and permutations. 
The most general $N=2$ lagrangian in 5d is usually written in terms of the quantity ${\cal V}={\frac{1}{6}}C_{IJK}X^{I}X^{J}X^{K}=X^1 X^2 X^2 = 1$, see more details in e.g.\ \cite{Klemm:2000nj}. Using this language we can reproduce the lagrangian \eqref{5dlagr} after identifying the two sections $X^1(z), X^2(z)$ as follows:
\begin{equation}
 X^1 = z, \qquad X^2 = \sqrt{z}^{-1}\ .
\end{equation}
This formalism is useful later when we consider the 5d to 4d reduction. Finally we observe that this lagrangian is also embeddable in the gauged $N=4^0$ theory of Romans \cite{Romans:1985ps} (see also \cite{Gunaydin:1984qu,Pernici:1985ju,Gunaydin:1985cu,Hristov:2013xza}).  

We now look more explicitly in the KK reduction along AdS and SS reduction along the sphere seperately and in each case show that the reduction preserves all or half of the supersymmetries, respectively. 

\subsection{KK on AdS$_3$} 
\label{subsec:31}
Performing a KK reduction leads us to an ungauged supergravity also in 5d, meaning that the bosonic lagrangian \eqref{5dlagr} is completed with the fermionic terms of ungauged $N=2$ supergravity. The near-horizon background we find here is obtained from the simple rules \eqref{redrules65} upon reducing along $\gamma$ in \eqref{ads_j}:
\begin{equation}
 ds^2_5 = L^2 \left(\frac{L}{2 J} \right)^{2/3} ( \frac{1}{4} {\rm d} s^2_{AdS_2}+{\rm d} s^2_{S^3}), \qquad z = \left(\frac{L}{2 J} \right)^{2/3}\ ,
\end{equation}
\begin{equation}
  F^1 =  \frac{L^2}{4 J}\ Vol (AdS_2), \qquad F^2 = - \frac{L}{2} Vol (AdS_2)\ .
\end{equation}
This is the fully BPS near-horizon geometry of the static BMPV black hole \cite{Breckenridge:1996is} in 5d $N=2$ ungauged supergravity, i.e.\ it preserves 8 real supercharges.

\subsection{SS on S$^3$} 
\label{subsec:32}
The SS reduction along the sphere introduces slightly different fermionic completion of the bosonic lagrangian \eqref{5dlagr}. Choosing the constant in the gravitino reduction ansatz \eqref{SSgravitino} to be $g_1$, we find the term $g_1 A^1$ in the gravitino covariant derivative, see \cite{Klemm:2000nj} for the general FI gauged $N=2$ supergravity in 5d. This leads us to a flat gauged $N=2$ theory that is equivalent with the case of $N=4^0$ supergravity also at fermionic level for abelian solutions discussed here. We find the following 5d solution when we reduce along the direction $\psi$ in \eqref{s_k}: 
\begin{equation}
 ds^2_5 = L^2 \left(\frac{L}{2 K} \right)^{2/3} ( {\rm d} s^2_{AdS_{3}}+ \frac{1}{4} {\rm d} s^2_{S^2}), \qquad z = \left(\frac{L}{2 K} \right)^{2/3}\ ,
\end{equation}
\begin{equation}
  F^1 = \frac{L^2}{4 \sigma K}\ Vol (S^2), \qquad F^2 = -\frac{L}{2} Vol (S^2)\ .
\end{equation}
Checking our answers with the known black string near-horizon geometries in 5d, we find an exact match with the solutions in section 4.2.2 in \cite{Hristov:2013xza}. It is also easy to check for supersymmetry by again using the results of \cite{Hristov:2013xza} in section 5.2. We find one extra condition on the above solution that fixes the arbitrary reduction constant $g_1$ in terms of the magnetic charge of the graviphoton,
\begin{equation}
  g_1 p^1 = -1, \qquad \rightarrow \quad g_1 = - \frac{4 \sigma K}{L^2}\ .
\end{equation}
Fixing $g_1$ ensures that the black string attractor obtained from the SS reduction is quarter-BPS in the $N=4^0$ theory or half-BPS in the flat gauged $N=2$ supergravity, i.e.\ it preserves 4 real supercharges \cite{Hristov:2013xza}. BPS-ness is achieved when the product of the two magnetic charges is negative:
\begin{equation}
p^1 p^2 = -\frac{L^3}{8 \sigma K} < 0, 
\end{equation}
which requires $\sigma = +1$, and therefore $g_1$ to be negative. When $\sigma = -1$ we have $p^1 p^2 > 0$ and the solution is non-BPS in the flat gauged supergravity and fully BPS in the ungauged supergravity that we retrieve upon setting $g_1 = 0$, see \cite{Hristov:2013xza}.	  

\section{5D to 4D} 
\label{sec:5to4}
We now perform the second step in the reduction, starting from the bosonic lagrangian of 5d supergravity and reducing it to 4d in the two separate cases that we obtained in the previous section. The 5d to 4d reduction is well-studied and we decide to be brief in the generalities, referring to \cite{Andrianopoli:2004im,Behrndt:2005he,Looyestijn:2010pb} for all details. The action \eqref{5dlagr} upon reduction gives the bosonic content of 4d $N=2$ ungauged supergravity with 2 extra vectormultiplets. The field content is therefore the metric $g_{\mu \nu}$, 3 gauge fields $A^{0,1,2}_{\mu}$, and 2 complex scalars $z^1, z^2$. The full ansatz for reducing the bosonic fields from 5 to 4 dimensions is the following:
\begin{equation}
 {\rm d} s^2_5 = e^{2 \phi} {\rm d} s^2_4 + e^{-4 \phi} ({\rm d} x_5 + A^0_4)^2\ ,
\end{equation}
\begin{equation}
 A^I_5 = A^I_4 + {\rm Re} z^I ({\rm d} \gamma + A^0_4)\ ,
\end{equation}
\begin{equation}
 X^I_5 = 2 e^{2 \phi}\ {\rm Im} z^I\ ,
\end{equation}
where the 4d fields are already in the standard 4d $N=2$ conventions and on the left hand side we have the 5-dimensional fields of \eqref{5dlagr}.
The resulting supergravity in four dimensions is defined by the prepotential (derived from the 5d coefficients $C_{I J K}$)
\begin{equation}\label{prepot}
 F = \frac{C_{I J K} X^I X^J X^K}{6 X^0} = \frac{X^1 (X^2)^2}{X^0}\ ,
\end{equation}
where the holomorphic sections $X^{\Lambda}$ are related to the scalars via $z^I = X^I/X^0$. For this prepotential and imaginary scalars, the period matrix of special geometry \cite{Andrianopoli:1996cm} is purely imaginary and diagonal: 
\begin{align}
I_{00}= & - {\rm Im} z^1 ({\rm Im} z^2)^2\\
I_{11}= & - \frac{({\rm Im} z^2)^2}{{\rm Im} z^1 } \\
I_{22}= & - 2 {\rm Im} z^1\ .
\end{align}
We need these values explicitly in order to find the conserved electric charges 
\begin{equation}
 Q_{\Lambda} = I_{\Lambda \Sigma} q^{\Sigma}\ ,
\end{equation}
which can be obtained from the original fields strengths $F^{\Lambda} = q^{\Lambda}\ Vol (AdS_2)$ via the period matrix $I$. 

\subsection{SS on S$^3$} 
\label{subsec:41}
Taking first the fully BPS black hole background in 5d from section \ref{subsec:31}, we now perform an SS reduction over the $\phi$ direction of S$^3$. This leads to gauging of the gravitino with the KK gauge field $g_0 A^0$ with the expected arbitrary constant $g_0$ from the SS reduction. We are then left with a flat FI gauged supergravity with prepotential \eqref{prepot} and only nonvanishing FI parameter $g_0$ (i.e.\ the other gauge fields are not used in the gauging, $g_1 = g_2 = 0$). Using the reduction rules above, 
we identify $$e^{-2 \phi} = \frac{L}{2 K} \left(\frac{L}{2 J} \right)^{1/3}$$ and find the following 4d background:
\begin{equation} \label{4dks}
 ds^2_4 = \frac{L^4}{16 J K} ( {\rm d} s^2_{AdS_2}+{\rm d} s^2_{S^2}), \qquad z^1 = \frac{i L^2}{8 J K}, \quad z^2 = \frac{i L}{4 K}\ , 
\end{equation}
\begin{equation} 
  F^0 =  \sigma K\ Vol (S^2), \quad F^1 = \frac{L^2}{4 J}\ Vol (AdS_2), \quad F^2 = -\frac{L}{2}\ Vol (AdS_2)\ .
\end{equation}
This is the near-horizon solution of a black hole with two electric and one magnetic charge,
\begin{equation}\label{4dksg}
 p^0 = \sigma K\ , \qquad Q_1 = - \frac{L^2}{8 K} \ , \qquad Q_2 = \frac{L^3}{8 J K}\ . 
\end{equation}
Looking at the supersymmetry properties of such class of solutions, analyzed in \cite{Hristov:2012nu}, we again observe that there is one condition fixing the constant $g_0$ in terms of the magnetic charge,
\begin{equation}
 g_0 p^0 = -1, \qquad \rightarrow \quad g_0 = - \frac{1}{\sigma K}\ .
\end{equation}
This guarantees that preservation of 4 supercharges in the flat gauged $N=2$ supergravity in 4d, provided that the quartic invarian $I_4$ is negative:  \begin{equation}
I_4 = -2 p^0 Q_1 (Q_2)^2 = \sigma \left( \frac{L^4}{16 J K} \right)^2\ .
\end{equation}
Unlike the SS reduction of section \eqref{subsec:32}, the BPS condition $I_4 < 0$ now translates into $\sigma = -1$. Vice versa, when $\sigma = +1$, the solution is fully BPS in ungauged supergravity, where $g_0 = 0$.

\subsection{KK on AdS$_3$} 
\label{subsec:42}
In this case we start from already gauged fermionic completion of the 5d lagrangian \eqref{5dlagr}, but due to the KK reduction to 4d do not add extra gauging of the gravitino, therefore having the already fixed $g_1$ and leaving $g_0 = g_2 = 0$. This is another type of flat FI gauged supergravity with the same bosonic lagrangian as in the case above. The solution following from the reduction rules along $\gamma$ in AdS$_{3}$ is very similar,
\begin{equation} \label{4dsk}
 ds^2_4 = \frac{L^4}{16 J K} ( {\rm d} s^2_{AdS_2}+{\rm d} s^2_{S^2}), \qquad z^1 = \frac{i L^2}{8 J K}, \quad z^2 = \frac{i L}{ 4 J}\ , 
\end{equation}
\begin{equation}
  F^0 = J\ Vol (AdS_2), \quad F^1 = \sigma \frac{L^2}{4 K}\ Vol (S^2), \quad F^2 = -\frac{L}{2}\ Vol (S^2)\ .
\end{equation}
This is the near-horizon solution of a black hole with two magnetic and one electric charge,
\begin{equation}\label{4dskg}
 Q_0 = - \frac{ L^4}{128 K J^2}\ , \qquad p^1 = \sigma \frac{L^2}{4 K}\ , \qquad p^2 = - \frac{L}{2}\ . 
\end{equation}
We find that the quartic invariant $I_4$ for this solution is:
\begin{equation}
I_4 = 8 Q_0 p^1 (p^2)^2 = - \sigma \left( \frac{L^4}{16 J K} \right)^2\ .
\end{equation}
Exactly as in the case above, this is a half-BPS near-horizon solution (BPS'ness guaranteed by the condition $g_1 p^1 = -1$ that we already satisfied in 5d). Just like the corresponding 5d attractor upon which we reduced, the BPS constraint $I_4 < 0$ again gives $ \sigma = +1$. The opposite choice, $\sigma=-1$, leads to a fully BPS solution in ungauged supergravity for $g_1=0$.

\section{Commutator between KK-SS and SS-KK} 
\label{sec:commutator}
In the previous sections we showed how to obtain the half BPS attractor AdS$_2 \times $S$^2$ in four dimensions by reducing the fully BPS black string near horizon geometry AdS$_3 \times $S$^3$. We followed two different paths where the two operations of KK on AdS$_3$ and SS on S$^3$ are interchanged. We can now address the question about the commutator between these two operations, and the answer can be found after the use of a duality transformation.

$N=2\ D=4$ supergravity has symplectic transformations as electromagnetic duality group so that the fields can be organized into symplectic vectors. The vector describing the scalar degrees of freedom is:
\begin{equation}
\Omega = \left( \begin{array}{c}
      X^{\Lambda} \\
      F_{\Lambda}\end{array} \right) \ ,\ 
\end{equation}
where $F_{\Lambda}$ is defined via the prepotential as $F_{\Lambda} = \partial F/\partial X^{\Lambda}$.
A second symplectic vector is needed to describe the gauge degrees of freedom and contains electric and magnetic charges:
\begin{equation}
\Gamma = \left( \begin{array}{c}
      p^{\Lambda} \\
      Q_{\Lambda}\end{array} \right)\ .
\end{equation}
It is also possibile to define a symplectic vector containing the FI parameters: 
\begin{equation}
G = \left( \begin{array}{c}
      g^{\Lambda} \\
      g_{\Lambda}\end{array} \right)\ ,\ 
\end{equation}
where $g^{\Lambda}$ and $g_{\Lambda}$ correspond to electric and magnetic gaugings respectively (see \cite{deWit:2011gk} for more details on magnetic gaugings).

We consider a symplectic transformation of the type:
\begin{equation} \label{duality}
M =
 \begin{pmatrix}
  0 & 0 & 0 & a & 0 & 0 \\
  0 & 0 & 0 & 0 & b & 0 \\
  0 & 0 & 0 & 0 & 0 & c \\
  - a^{-1} & 0 & 0 & 0 & 0 & 0 \\
  0 & - b^{-1} & 0 & 0 & 0 & 0 \\
  0 & 0 & - c^{-1} & 0 & 0 & 0 
  \end{pmatrix}
\end{equation}
This particular duality transformation is not only a symmetry of the equations of motion, but also a symmetry of the bosonic lagrangian. It leaves the prepopential \eqref{prepot} invariant, provided that
\begin{equation}
 4 b = \frac{a}{c^2}
\end{equation}
is satisfied. This is exactly the type of transformation that we need in order to match the two BPS attractors. We start from the scalar fields, that transform under \eqref{duality} as: 
\begin{equation}
(z^{1})' = -\frac{a^{-1} b}{z^1} \ ,\ (z^{2})' = -\frac{2 a^{-1} c}{z^2}\ .
\end{equation}
The two sets of scalar fields in \eqref{4dks}, \eqref{4dsk} rotate into each other if we set the three parameters to be:
\begin{equation}
a = \pm \frac{128 J^2 K^2}{L^4} \ , \ b = \pm 2 \ ,\ c = \pm \frac{4 J K}{L^2}\ .
\end{equation}
Moving to the gauge sector we can write the two sets of charges \eqref{4dksg}, \eqref{4dskg} as symplectic vectors:
\begin{equation}
\Gamma = \left( \begin{array}{c}
       \sigma K  \\ 0 \\ 0 \\ 0 \\
      - \frac{L^2}{8 K}  \\
      \frac{L^3}{8 J K}
      \end{array} \right) \ ,\ 
      \Gamma '= \left( \begin{array}{c}
      0 \\ \frac{L^2}{4 \sigma K}  \\ -\frac{L}{2} \\
      - \frac{  L^4}{128 K J^2} \\ 0 \\ 0
      \end{array} \right) \ .
\end{equation}
These two vectors fail to transform into each other under duality only for a sign of one charge. This is where the equivalence between the two solutions breaks down, as expected since we showed that the two 4d backgrounds are BPS for opposite values of $\sigma$. The two BPS solutions are not exactly equivalent under duality, since an extra flip in the sign of the quartic invariant $I_4 \rightarrow - I_4$ is needed. 

Let us now consider the FI parameters:
\begin{equation}
G = \left( \begin{array}{c}
0 \\0 \\0\\  - \frac{1}{\sigma K} \\0 \\0       
\end{array} \right)
\ ,\ G ' = \left( \begin{array}{c}
0 \\0 \\0\\ 0 \\- \frac{4 \sigma K}{L^2} \\0      
\end{array} \right) \ .
\end{equation}
It is clear that the symplectic rotation \eqref{duality} does not map these two vectors, since it transforms electric gaugings into magnetic gaugings and vice versa, i.e.\ it transforms $g_0$ into $g^0$ and $g^1$ into $g_1$. Here lies the nontrivial part of the commutator between the two operations KK on AdS$_3$ and SS on S$^3$, which amounts to interchanging electric and magnetic gaugings.

\section{The dual CFT picture and black hole/string microscopics} 
\label{sec:dualCFT}

In the previous sections we were able to give a clear string theory interpretation to the supersymmetric near horizon geometries of non-BPS black holes \cite{Hristov:2012nu} and non-BPS black strings \cite{Hristov:2013xza}. This we achieved starting from the 6d black string attractor AdS$_3 \times$ S$^3$, which corresponds to the near horizon geometry of the D1-D5 system on K3 or T$^4$, and reducing it to lower dimensions. We now make use of AdS/CFT techniques to understand the properties of the field theories that are dual to the 5d and 4d solutions we presented.

Our starting point is the 2d $N = (4, 4)$ SCFT \cite{Strominger:1996sh} on $\mathbb{R}^{1,1}$ describing the D1-D5 system degrees of freedom, dual to the 6d black string background. Performing a KK reduction to the 5d black hole of section \ref{subsec:31} is equivalent to putting the $(4, 4)$ theory on a circle preserving supersymmetry. This is by now a well-established fact that enabled the microscopic entropy counting of the black holes in 5d \cite{Strominger:1996sh}. 

What about the meaning of the reduction to the 5d black string of section \ref{subsec:32} from a field theory point of view? We know that the operation of SS reduction along S$^3$ breaks the $SU(2)\times SU(2)$ symmetry group of the sphere to $U(1) \times SU(2)$ in the lower dimension (keeping the $U(1)$ subgroup of $SU(2)$ due to the reduction ansatz). This is also true for the full geometry where in general supersymmetry is broken. On the horizon we restore the supersymmetry group to $SU(1,1|1) \times SU(1,1) \times SU(2)$, i.e.\ with respect to the original attractor before the reduction we break the R-symmetry from $SU(2)_R$ to $U(1)_R$ and keep the $SU(2)_L$ only as a global symmetry. This is crucial to understand the properties of the dual field theory that follow from the AdS/CFT dictionary. We know that the operation of SS reduction breaks fully the supersymmetry of the D1-D5 system and leads to a $N=(0,0)$ theory that then flows in the IR to a $N = (0,2)$ SCFT on $\mathbb{R}^{1,1}$. This means that we have completely broken supersymmetry in the left-moving sector, and broken the R-symmetry to $U(1)_R$ in the right moving sector in the infrared\footnote{Note that even the straightforward KK reduction from S$^3$ to S$^2$ breaks completely the supersymmetry on the left-moving sector, but keeps the $SU(2)_R$. This does not lead to a change in the underlying degrees of freedom of the D1-D5 system.}. 

It therefore seems natural to expect that the operation of SS reduction projects out the states in the D1-D5 field theory that are charged under the original $SU(2)_R$ R-symmetry, therefore finding in principle a smaller number of massless fields in the $N = (0, 0)$ theory as compared to its parent $N = (4, 4)$. The consequent flow to the $N=(0,2)$ supersymmetric fixed point should also change accordingly and due to the lack of supersymmetry along the RG flow it seems unclear how exactly this IR theory can be defined. However, as we show later in this section, the value of the central charge does not change, and in our naive picture this happens for a simple reason: the central charge in the D1-D5 system (see a detailed discussion in section 5.3.1 of \cite{Aharony:1999ti} and in \cite{Hassan:1997ai}) counts the number of massless hypermultiplets in the large charge supergravity limit. The left moving hypermultiplet fields are inert under the $SU(2)_R$, so they remain unaffected by the SS reduction and therefore we get the same central charge in the resulting $N = (0,2)$ theory\footnote{Interestingly, one can find the analogous reasoning on the gravity side by the argument that BPS states making up the black hole entropy need to be rotationally (i.e. $SU(2)$) invariant \cite{Dabholkar:2010rm}.}.

Going down to the 4d near-horizon geometries of sections \ref{subsec:41} and \ref{subsec:42} we find the same field theory description, given by the new $N = (0, 2)$ field theory on a spatial circle. One should then be able to extract the value of the Bekenstein-Hawking entropy of the black holes from their dual description. This is indeed the case, as we show now.

Starting from 6d, we know from \cite{Strominger:1996sh} that the value of the central charge for the D1-D5 field theory is\footnote{Here and in what follows we have defined $A_2 $, $A_3 $ to be the integral of the respective volume element:
\begin{equation}
A_2 (L) = L^2 \int Vol (S^2) = 4 \pi L^2\ , \qquad A_3 (L) = L^3 \int Vol (S^3) = 2 \pi^2 L^3\ .
\end{equation}}:
\begin{equation}\label{centralcharge}
 c = \frac{3 R_{AdS_{3}}}{2 G_3} = \frac{3 L A_3 (L)}{2 G_6} = \frac{3 \pi^2 L^4}{G_6} \ .
\end{equation}
This result follows from the standard AdS/CFT dictionary, but as already discussed above it was derived independently on the field theory side in \cite{Strominger:1996sh} (see again \cite{Aharony:1999ti}). Going down along AdS$_3$ puts the theory on a circle with momentum (see e.g.\ \cite{Kraus:2006wn})
\begin{equation}
 L_0 - \frac{c}{24} = \rho_+^2 \frac{R_{AdS_{3}}}{4 G_3} = \frac{R_{AdS_{3}}}{16 J^2 G_3} = \frac{L A_3 (L)}{16 J^2 G_6} = \frac{\pi^2 L^4}{8 J^2 G_6}\ ,
\end{equation}
where $\rho_+$ is the horizon radius of BTZ in standard notation (see e.g.\ \cite{Hristov:2014eza}) that translates for us into $\rho_+ = (2 J)^{-1}$. The Cardy formula therefore leads to 
\begin{equation}\label{cardyform}
 S_{Cardy} = 2 \pi \sqrt{\frac{c}{6} \left(L_0 - \frac{c}{24} \right)} = \frac{\pi R_{AdS_{3}}}{4 J G_3} = \frac{\pi^3 L^4}{2 J G_6}\ .
\end{equation}
If one is careful about dimensional analysis, this formula matches exactly with the Bekenstein-Hawking entropy formula
\begin{equation}
 S_{BH} = \frac{A_d}{4 G_{d+2}}\ ,
\end{equation}
for the black hole in 5d of section \ref{subsec:31} and both black holes in section \ref{sec:5to4}. We need the respective 5d and 4d Newton constants that can be derived from the (in this case more fundamental) 6d Newton constant,
\begin{equation}\label{newtonconst}
 \frac{1}{G_6} = \frac{1}{2 \pi G_5} = \frac{1}{4 \pi K G'_5} = \frac{1}{8 \pi^2 K G_4}\ , 
\end{equation}
where there are two different normalizations of the Newton constant in 5d, depending on whether we first reduce along AdS ($G_5$) or along the sphere ($G'_5$). Going further to 4d we obtain that $G_4 = G'_4$ as the two reduction paths converge.

More explicitly, we find the 5d black hole to have
\begin{equation}
 S^5_{BH} = \frac{A_3 (L (L/2 J)^{1/3})}{4 G_5} = \frac{\pi^2 L^4}{4 J G_5}\ ,
\end{equation}
while for the 4d black holes 
\begin{equation}
 S^4_{BH} = \frac{A_2 (L^2/(4 \sqrt{J K}))}{4 G_4} = \frac{\pi L^4}{16 J K G_4}\ .
\end{equation}
These two expressions are equal to each other and also equal to the Cardy formula \eqref{cardyform} of the dual field theory upon imposing \eqref{newtonconst}.

We also recover correctly the central charge of the 5d black string, given by the AdS/CFT formula \cite{Brown:1986nw}
\begin{equation}
 c = \frac{3 R_{AdS_{3}}}{2 G'_3} = \frac{3 L (L/2 K)^{1/3} A_2 (L/2\ (L/2 K)^{1/3})}{2 G'_5} = \frac{3 \pi L^4}{4 K G'_5}\ ,
\end{equation}
which matches \eqref{centralcharge} upon the identification \eqref{newtonconst}. 

This proves our expectation that the central charge of the $N=(4,4)$ theory remains unchanged even after projecting out states charged under $SU(2)_R$ to end up with the new $N=(0,2)$ theory. 

\section{Concluding remarks} 
\label{sec:conclusion}
To briefly summarize our results, we followed two different dimensional reduction paths from the black string attractor AdS$_3 \times$S$^3$ to AdS$_2 \times$S$^2$. We looked explicitly only at the attractors, but it is straightforward to write the reduction for the full flows to asymptotic Minkowski as done in \cite{Breckenridge:1996is,Behrndt:2005he}. Our main point was to show supersymmetry on the horizon, for which zooming in on the attractor geometry was enough.

Preservation of supersymmetry is the reason to claim that our example is a genuine case of dual microscopic description. The usual check that we performed in the matching between the Cardy and Bekenstein-Hawking entropy was already known to work for any extremal black hole \cite{Strominger:1997eq}. The fact that we keep supersymmetry on the horizon is the extra ingredient that keeps the dual description under control. One can therefore hope to match the macroscopic and microscopic entropy formulas also after taking into account quantum corrections. This would however require a more detailed understanding of the resulting $N=(0,2)$ 2d SCFT, as well as looking at higher derivative corrections on the gravity side, see \cite{Sen:2014aja,Baggio:2014hua} and references therein.

Apart from this main purpose of our work, we also showed that the operations of KK reduction on AdS$_3$ and SS reduction S$^3$ and their exchange do not commute. After the duality transformation we constructed in section \ref{sec:commutator} it turned out that if one path gives a standard electric gauging in the 4d supergravity, the other one can be seen as magnetic gauging. The two different reduction paths also lead us to different topological sectors of black hole solutions, exchanging the quartic invariant $I_4$ with its opposite value $-I_4$.

\section*{Acknowledgments}
We would like to thank S.\ Katmadas, A.\ Tomasiello, S.\ Vandoren, and A.\ Zaffaroni for enlightening discussions. We are supported in part by INFN, by the MIUR-FIRB grant RBFR10QS5J ``String Theory and Fundamental Interactions'', and by the MIUR-PRIN contract 2009-KHZKRX.

\providecommand{\href}[2]{#2}
\end{document}